\documentclass[12pt]{article}

\usepackage{amsmath,amsfonts,amssymb,a4}
\usepackage[latin1]{inputenc}
\usepackage{verbatim}

\usepackage{color}

\def\bB {{\mathbb{B}}}
\def\bT {{\mathbb{T}}}
\def\bS {{\mathbb{S}}}
\def\bR {{\mathbb{R}}}
\def\bN {{\mathbb{N}}}
\def\bC {{\mathbb{C}}}

\def\bO {{\mathbb{O}}}
\def\bZ {{\mathbb{Z}}}

\def\bS {{\mathbb{S}}}
\def\bU {{\mathbb{U}}}
\def\bW {{\mathbb W}}

\def\cO {{\mathcal O}}

\def\bbO {{\mathbf{O}}}

\def\Di {\displaystyle}

\def\MR {\mathrm }

\def\sp {\mathrm {sp}}

\def\No {\noindent}
\def\Bar {\, |\, }

\makeatletter
\@addtoreset{equation}{section}

\makeatother

\newtheorem{theorem}{Theorem}[section]

\newtheorem{proposition}[theorem]{Proposition}

\def\demo {\noindent {\it Proof. }}
\def\fin {\square}

\title{Semiclassical analysis for a Schr\"odinger operator
with a U(2) artificial gauge: the periodic case}
\author{A. Morame$^1$ and F. Truc$^2$}
\date{June 16, 2014}

\begin{document}

\bibliographystyle{plain}

\maketitle

\vskip 0.5cm

\begin{center}
{\it {$^{1}$ Universit\'e de Nantes,
Facult\'e des Sciences,  Dpt. Math\'ematiques, \\
UMR 6629 du CNRS, B.P. 99208, 44322 Nantes Cedex 3, (FRANCE), \\
E.Mail: Abderemane.Morame@univ-nantes.fr}}\\
{\it {$^{2}$ Universit\'e de Grenoble I, Institut Fourier,\\
            UMR 5582 CNRS-UJF,
            B.P. 74,\\
 38402 St Martin d'H\`eres Cedex, (France), \\
E.Mail: Francoise.Truc@ujf-grenoble.fr }}
\end{center}



\begin{abstract}
We consider a Schr\"odinger operator with a Hermitian
2x2 matrix-valued potential which is lattice periodic and can be
diagonalized smoothly
on the whole $R^n.$
In the case of  potential taking its minimum only on the lattice, we
prove that the well-known semiclassical asymptotic of first
band spectrum for a scalar potential remains valid for our model.
\end{abstract}

\textbf{Keywords~:} semiclassical asymptotic, spectrum, eigenvalues,
Schrodinger, periodic potential, BKW method, width of the first band,
magnetic field.

\textbf{AMS MSC 2000~:} 35J10, 35P15, 47A10, 81Q10, 81Q20.

\tableofcontents

\section{Introduction}
Schr\"odinger operators with periodic matrix-valued potentials appear in
many models in physics.
Such models  have been used recently to describe the motion of an atom in
optical fields (\cite{Co},   \cite{Co-Da}, \cite{Da-al}), see also
\cite{Ca-Yu}.
The aim of this paper is to investigate their spectral properties using
semiclassical analysis. We focus on the first spectral band and
assume that the potential has a non degenerate minimum.
The Schr\"odinger operators with a non-Abelian gauge potential  are 
Hamiltonian operators
on $L^2(\bR^n; \bC^m)$
of the  following  form~:
\begin{equation}\label{type}
\begin{array}{c}
  H^h\,  =\, h^2 \sum_{k=1}^{n}(D_{x_k}I -A_k)^2 + V + hQ +h^2R=P^h(x,hD)\; .
 \end{array}
 \end{equation}
The classical symbol of $P^h(x,hD),\ P^h(x,\xi ),$ for $ (x,\xi )\in
\bR^n\times \bR^n,$ is given by
\begin{equation}\label{typeP}
  \quad P^h(x,\xi )=\sum_{k=1}^{n}\left \{ (\xi_kI -hA_k(x))^2
  +ih^2\partial_{x_k}A_k(x)\right\}  + V(x) + hQ(x) +h^2R(x)\; ,
\end{equation}
 $I$ is the identity $m\times m$ matrix, $V,\ Q,\ R$ and the $A_k$
 are hermitian $m\times m$ matrix with smooth coefficients
 and $\Gamma $ periodic:
 \begin{equation}\label{coeff}
 \left .
\begin{array}{l}
A_k=(a_{k,ij}(x))_{1\leq i, j\leq m},\quad
  V=(v_{ij}(x))_{1\leq i, j\leq m},\\
  Q=(q_{ij}(x))_{1\leq i, j\leq m},\quad
  R=(r_{ij}(x))_{1\leq i, j\leq m},\\
  a_{k, ij},\
  v_{ij},\ q_{ij},\ r_{ij} \in C^\infty (\bR^n;\bC) ,\\
  \overline{a_{k, ji}}=a_{k,ij},\
  \overline{v_{ji}}=v_{ij},\ \overline{q_{ji}}=q_{ij},
  \ \overline{r_{ji}}=r_{ij}\\
   a_{k, ij}(x-\gamma )=a_{k,ij}(x),\
  v_{ij}(x-\gamma )=v_{ij}(x),\\
   q_{ij}(x-\gamma )=q_{ij}(x) \ \MR{and} \ r_{ij}(x-\gamma )=r_{ij}(x)\
  \forall \gamma \in \Gamma \; ;
 \end{array}
 \right \}
 \end{equation}
 $\Gamma $ is a lattice of $\Di \bR^n,\ \Gamma =\{ \sum_{k=1}^{n}  m_k
\beta_k ; \ m_k \in \bZ \} ,$\\
 $ \beta_1, \beta_2,\ldots , \beta_n\, \in \bR^n$ form
 a basis, $det( \beta_1, \beta_2,\ldots , \beta_n)\neq 0.$\\
  We use the notation  $D = (D_{x_1},\ldots , D_{x_n})$
  where $ D_{x_k}=-i\partial_{x_k}, \ k=1...n,$\\
  so $D^2=-\Delta $ is the Laplacian
  operator on $L^2(\bR^n).$

The dual basis $\{ \beta_1^\star , \ldots ,\beta_n^\star \} $ of the
reciprocal lattice $\Gamma^\star ,$
is
the basis of $\bR^n$
defined by the relations
\[ \beta_j^\star .\beta_k =2\pi \delta_{jk}\, :\quad \Gamma^\star =\{
\sum_{k=1}^{n}  m_k \beta_k^\star  ; \ m_k \in \bZ \}  .
\]
 The fundamental cell, the Wigner-Seitz cell,
 \[ \bW^n = \{ \sum_{k=1}^{n} x_k\beta_k\, ;\ x_k\in ]-\frac{1}{2},
 \frac{1}{2}[\} \, ,\]
 will be identified with the n-dimensional torus $\bT^n=\bR^n/\Gamma $ and
the dual cell,
the Brillouin zone,  is defined by
\[ \bB^n = \{ \sum_{k=1}^{n}  \theta_k \beta_k^\star  ; \theta_k \in
]-\frac{1}{2},-\frac{1}{2}[\}\ .
\]
We will identify $L^2(\bT^n; \bC^m) $ with $\Gamma $ periodic functions
of $L^{2}_{loc}(\bR^n; \bC^m) $
provided with the norm of $L^2(\bW^n; \bC^m) .$
In the same way the Sobolev space $W^k(\bT^n; \bC^m) ,$ with
$k\in \bN ,$
may be identified with $\Gamma $ periodic functions
of $W^{k}_{loc}(\bR^n; \bC^m) $
provided with the norm of $W^k(\bW^n; \bC^m) .$

By Floquet theory, (see \cite{Ea} or \cite{Re-Si} ),  we have
\[ H^h = \int_{\bB^n }^\oplus H^{h,\theta} d\theta\ ,\]
with $ H^{h,\theta}$  the partial differential operator
$ P_h(x, h(D-\theta))$ on $L^2(\bT^n; \bC^m) .$ \\
The ellipticity of $ P_h(x, h(D-\theta))$ implies that the spectrum of
 $ H^{h,\theta}$ is discrete
 \begin{equation}\label{spHht}
 \sp(H^{h,\theta})=\{ \lambda_j^{h,\theta};\ j\in \bN^\star \} , \
\lambda_1^{h,\theta}\leq\lambda_2^{h,\theta}\leq \ldots \leq
\lambda_j^{h,\theta}\leq\lambda_{j+1}^{h,\theta}\leq \ldots
\end{equation}
each $\lambda_j^{h,\theta}$ is an eigenvalue of finite multiplicity and
each eigenvalue  is repeated according to its multiplicity. \\
(When $m=1$ and $\Di V=Q=R=A_k=0,\ \
 (\frac{1}{\sqrt{|\bT^n|}}e^{i\omega .x})_{\omega \in \Gamma^\star }$
is the Hilbert basis of $L^2(\bT^n)$
which is composed  of eigenfunctions
of $h^2(D-\theta )^2).$

 The Floquet theory guarantees that
 \begin{equation}\label{FloqS}
 \sp (H^h)=\bigcup_{\theta \in \bB^n} \ \sp(H^{h,\theta})
=\bigcup_{j=1}^{\infty} b_j^h\, ,
\end{equation}
where $b_j^h$ denotes the $j$-th band $b_j^h= \{\lambda_j^{h,\theta},
\theta \in\bB^n\}.$

In the sequel $h_0$ will be a non negative small constant, $h$ will be in
$]0,h_0[,$ and
any non negative constant which doesn't   depend on $h$ will
invariably  be
denoted  by $C.$

\section{Preliminary: the artificial gauge model}

We will be interested in the  model of artificial gauge  considered in 
\cite{Co}, \cite{Co-Da}    and \cite{Da-al}
\begin{equation}\label{Type1}
\left .
\begin{array}{l}
m=2,\ V=vI+W,\ A_k=Q=R=0,\ \forall k,\\
W= w . \sigma ,\ \MR{with}\  w =(w_1, w_2, w_3) , \ v\ \MR{and\ the}\ w_j\
\MR{are\ in}\  C^\infty(\bR^n;\bR),
\end{array} \right \}
\end{equation}
 we denote $\sigma = (\sigma_1, \sigma_2, \sigma_3),$ where the $\sigma_j$
are the Pauli matrices.

 Let us remark that
\begin{equation}\label{carre}
V= vI+W,\quad W=w.\sigma ,\quad W^2= |w|^2 I \ .
\end{equation}

In the sequel we will assume that
\begin{equation}\label{H1}
\left .
\begin{array}{l}
|w|\, >\, 0\\
v(x)-|w(x)|\
 \MR{has\ a\ unique\ non\ degenerate\ minimum\ on} \ \bT^n.
 \end{array}
 \right \}
\end{equation}

Due to the invariance of the Laplacian by translation and by the action of
$\bO (n),$
  we can assume, up to a composition by a translation of the potentials, that
  \begin{equation}\label{H1b}
  \left .
  \begin{array}{l}
v(\gamma )-|w(\gamma)|<v(x)-|w(x)|,\ \forall x\in \bR^n\setminus \Gamma
 \ \MR{and}\ \forall \gamma \in \Gamma ,\\
 v(x)-|w(x)|=E_0+\displaystyle{\sum_{k=1}^{n} }\tau_k^2 x_k^2 +
\mathbf{O}(|x|^3),\ \MR{as}\ |x|\to 0,
 \end{array}
 \right \}
 \end{equation}
$ (\tau_k>0,\ \forall k). $


There exists   $U\in \bU (2),$ ( a unitary $2\times 2$   matrix),   such that
\begin{equation}\label{VDtype1}
U^\star VU = \widetilde{V} = \left[\begin{array}{ccc} v-|w| & 0\\
0 & v+ |w|
\end{array} \right] \ .
\end{equation}
As $|w|$ never vanishes, $U=U(x)$ can be chosen smooth and
$\Gamma $ periodic:
\[ U=\left ( \begin{array}{cc} u_{11} & u_{12}\\
u_{21} & u_{22} \end{array} \right )
\, \in \,  C^\infty (\bT^n; \bU (2));\]
for example
$\Di  u_{11}=\frac{1}{2\sqrt{|w|(|w| - \MR{Re}((w_1 + i
w_2)e^{-i\theta}))}} (w_3 -|w| +e^{i\theta}(w_1-i w_2)) ,$

$\Di
\  u_{21}=\frac{1}{2\sqrt{|w|(|w| - \MR{Re}(w_1 + i w_2)e^{-i\theta}))}}
(w_1 + i w_2 -e^{i\theta}(w_3 +|w|)),$\\
$\Di u_{12}= \overline{u_{21}},\ u_{22}= -\overline{u_{11}}$ and
$\Di \theta  = \chi (\frac{w_2^2+w_3^2}{|w|^2}) \frac{\pi}{2},$\\
where $\chi (t)$ is a smooth function on the real line, $0\leq \chi
(t)\leq 1,$

$ \chi (t)=1$ when $|t|\leq 1/4$
and $\chi (t)=0$ when $|t|\geq 1/2.$
So
\begin{equation}\label{Utype1}
U =(\alpha ,\beta ,\rho ).\sigma +i\delta \sigma_0\, ,\quad
\MR{with}\  (\alpha , \beta , \rho , \delta )\in
C^\infty (\bT^n; \bS^3 );
\end{equation}
$\sigma_0$ is the $2\times 2$ identity matrix and $\bS^3$ is the unit
sphere of $\bR^4 .$

When $w_1+ i w_2 \neq 0$ or when $w_3<0,$ one can choose $U$ such that
$\delta =0$ by taking
$\Di  (\alpha ,\beta ,\rho )=\frac{1}{\sqrt{2|w|}}
(-\frac{w_1}{\sqrt{|w|-w_3}} ,
-\frac{w_2}{\sqrt{|w|-w_3}} , \sqrt{|w|-w_3}) .$

 Firstly let us expand the formula of the operator
\[ \widetilde{H}^h =  U^\star H^hU
 =h^2 D^2 I + U^\star VU
  -2i h^2\sum_{k=1}^{n}\left [ (U^\star  \partial_{x_k}U) D_{x_k} - h^2 
U^\star \partial_{x_k}^2 U
\right ]
\]
which can be rewritten as
\begin{equation}\label{pagauge}\widetilde{H}^h =  U^\star H^hU =
h^2 \sum_{k=1}^{n}(D_{x_k}I -A_k)^2 + \widetilde{V}  +h^2R
\, ,    \end{equation}
where $ A_k =i U^\star  \partial_{x_k}U :$
\begin{equation}\label{Atype1}
 A_k  =[(\partial_{x_k}\alpha , \partial_{x_k}\beta , \partial_{x_k}\rho )
\wedge (\alpha , \beta , \rho ) + (\delta \partial_{x_k} \alpha - \alpha
\partial_{x_k}\delta ,
\delta \partial_{x_k}\beta - \beta \partial_{x_k} \delta , \delta
\partial_{x_k} \rho
-\rho \partial_{x_k} \delta )].\sigma \, ,
\end{equation}
and
\begin{equation}\label{Rtype1}
R = \sum_{k=1}^{n} \left \{  (U^\star \partial_{x_k} U)^2 +
(\partial_{x_k} U^\star ).(\partial_{x_k} U)\right \} \, .
\end{equation}
So we can assume that $H^h$ is of the form (\ref{type}) with
$m=2,\ Q=0,\ A_k$ and $R$  given by (\ref{Atype1}) and
(\ref{Rtype1}), with $U$ defined by (\ref{Utype1}), and
$V=\widetilde{V}$ a diagonal matrix given by (\ref{VDtype1}).

\begin{theorem}\label{bands}
Under the above assumptions,  the first bands $ b_j^h, \ j=1, 2, \ldots ,$
of $H^h$  are concentrated around the value $ h\mu_j + E_0 \ j=1, 2,
\ldots ,$ in the sense that, there exist $ N_0>1$
and $h_0>0$ such that
\[ distance(h\mu_j + E_0, b_j^h) \leq C h^2, \quad
\forall j<N_0\ \MR{and}\ \forall h,\ 0 <h<h_0,
\]
where  $\Di \mu_j=\sum_{k=1}^{n} (2j_k+1)\tau_k,\ j_k\in \bN,$
the $\Di (\mu_\ell )_{\ell \in \bN^\star}$
is the increasing sequence of the eigenvalues of the harmonic oscillator
$\Di -\Delta + \sum_{k=1}^{n} \tau_k^2 x_k^2.$
\end{theorem}


\section{ Proof of Theorem~\ref{bands}}
\demo
According to  the above discussion, we can assume that
\begin{equation}\label{Hmodel1}  H^h =P^h(x, hD)\, ,\quad
\MR{with}\ P^h(x,hD)= \left( \begin{array}{cc}
P^{h}_{11}(x,hD) & P^{h}_{12}(x,hD) \\
P^{h}_{21}(x,hD)& P^{h}_{22}(x,hD)\end{array} \right )\,  ,
\end{equation}
with
\begin{equation}\label{Hmodel1b}
\left .
\begin{array}{l}
P^{h}_{11}(x,hD) =  h^2 (D - a_{.,11} (x))^2 \, +\, v(x)-|w(x)|\, +\,
h^2r_{11}(x)\\
P^{h}_{22}(x,hD) =  h^2 (D + a_{.,11} (x))^2 \, +\, v(x)+|w(x)|\, +\,
h^2r_{22}(x)\\
 P^{h}_{12}(x,hD) = - h^2 a_{. ,12}(x).(D + a_{.,11} (x)) -h^2 a_{.
,12}(x).(D -  a_{.,11} (x)) \\
\quad \quad \quad +\,  ih^2 div(a_{. ,12}(x)) \, +\, h^2r_{12}(x)\\
P^{h}_{21}(x,hD) = - h^2 a_{.,21}(x)(D - a_{.,11} (x))  -
h^2a_{.,21}(x).(D + a_{.,11} (x))\\
\quad \quad \quad
+\, ih^2 div(a_{.,21}(x)) \, +\, h^2r_{21}(x)\, .
\end{array}
\right \}  \end{equation}
$(D=(D_{x_1}, D_{x_2},\ldots , D_{x_n})$ and $a_{.,ij}(x)=(a_{1,ij}(x),
a_{2,ij}(x),\ldots , a_{n,ij}(x))\, .)$ \\
(We used that $ a_{.,22}=-a_{.,11}).$

Let us denote by  $H^{h,\theta}_{11}$ and $H^{h,\theta}_{22}$ the
    operators associated with \\
$P^{h}_{11}(x,h(D-\theta ))$ and $P^{h}_{22}(x,h(D-\theta ))$
on $L^2(\bT^n; \bC ).$\\
Then, if $\Di c_0= \min |w(x)|$ and $c_1=\max \| R(x)\| ,$
\[
\sp (H^{h,\theta}_{11})\, \subset \, [E_0-h^2c_1, +\infty [
\quad \MR{and} \quad \sp (H^{h,\theta}_{22})\, \subset \, [E_0-h^2c_1
+2c_0, +\infty [ .
\]
To prove the theorem  it is then enough to prove the proposition below.
\begin{proposition}\label{bandd} Let us consider a constant $c,\ 0<c<c_0.$
Then there exists $C_0 >0$ such that, for any
  $E^h \in  ]-\infty , E_0 +2c[ ,$
  we have
\begin{equation}\label{Ebandd}
\left . \begin{array}{l}
E^h \in \sp ( H^{h,\theta })\ \Rightarrow \
distance ( E^h, \sp( H^{h,\theta}_{11})\leq C_0 h^2\\
E^h \in \sp ( H^{h,\theta }_{11})\ \Rightarrow \
distance ( E^h, \sp( H^{h,\theta}))\leq C_0 h^2\, .
\end{array} \right \}
\end{equation}
\end{proposition}
\demo
For such $E^h,\ (H^{h,\theta}_{22}-E^h)^{-1}$ exists and, thanks to
semiclassical pseudodifferential calculus of
\cite{Ro} (see also \cite{Di-Sj} ), for $h_0>0$ small,
if $0<h<h_0$ then \\
 $\Di \| (H^{h,\theta}_{22}-E^h)^{-1}\|_{L^2(\bT^n)} +
 \|h(D-\theta ) (H^{h,\theta}_{22}-E^h)^{-1}\|_{L^2(\bT^n)} $\\
 $\Di
 + \| (H^{h,\theta}_{22}-E^h)^{-1} h(D-\theta )\|_{L^2(\bT^n)}
+ \| h(D-\theta ) (H^{h,\theta}_{22}-E^h)^{-1} h(D-\theta )\|_{L^2(\bT^n)}
 \leq C ,$\\
and then
\[
\| P_{12}^h(x, h(D-\theta ))
(H^{h,\theta}_{22}-E^h)^{-1}P_{21}^h(x, h(D-\theta ))
\|_{L^2(\bT^n)} \leq h^2C.
\]
 So if $E^h\in \sp ( H^{h,\theta }),$ then
$u^h=(u_1^h, u_2^h)\, \neq \, (0,0)$ is   an  eigenfunction of
$H^{h,\theta}$ associated with $E^h $ iff
\begin{equation}\label{system}
\left .
\begin{array}{l}
H^{h,\theta}_{11}u_1^h +P_{12}^h(x, h(D-\theta )) u_2^h = E^hu_1^h \\
u_2^h= - (H^{h,\theta}_{22}-E^hI)^{-1}P_{21}^h(x, h(D-\theta )) u_1^h \, .
\end{array}
\right \}  \end{equation}
In fact $E^h \in  ]-\infty , E_0 +c[ $
will be an eigenvalue of $H^{h,\theta}$ iff there exists\\
 $u_1^h$ in the Sobolev space $W^2(\bT^n ; \bC ), \ \| u_1^h
\|_{L^2(\bT^n)} \neq 0 ,  $
 such that
\[ H^{h,\theta}_{11}u_1^h  - P_{12}^h(x, h(D-\theta ))
(H^{h,\theta}_{22}-E^hI)^{-1}P_{21}^h(x, h(D-\theta )) u_1^h= E^hu_1^h
\, ,
\]
then we get the first part of  Proposition \ref{bandd}.

If $E^h$ is an eigenvalue of $H^{h,\theta}_{11}$ satisfying the assumption
of  Proposition \ref{bandd}, and $u_1^h$ an associated
eigenfunction,  then with \\
$\Di u^h= (u_1^h, - (H^{h,\theta}_{22}-E^h)^{-1}P_{21}^h(x, h(D-\theta ))
u_1^h),$  one has \\
\[  \begin{array}{l} \| (H^{h,\theta} - E^hI)u^h
\|_{L^2(\bT^n;\bC^2)} \\
= \| P_{12}^h(x, h(D-\theta ))
(H^{h,\theta}_{22}-E^h)^{-1}P_{21}^h(x, h(D-\theta ))u_1^h
\|_{L^2(\bT^n ;\bC )}\\
 \leq h^2C\|
u^h\|_{L^2(\bT^n;\bC^2)} \, ,
\end{array} \]
we get the second part of  Proposition \ref{bandd}.

Theorem \ref{bands} follows from Proposition \ref{bandd}
and \cite{Si-1},  \cite{Si-2}, \cite{He-Sj-1} and  \cite{He-Sj-2} results,
(see also \cite{He}),
which guarantee that
the sequence of eigenvalues of
$H^{h,\theta}_{11},\ (\lambda_j(H^{h,\theta}_{11}))_{j\in \bN^\star }$
satisfies
$\forall N_0>1,\ \exists h_0>0,\ C_0>0\ s.t.\
\forall h,\ 0<h<h_0$ and $\forall j\leq N_0\, , \quad
|\lambda_j(H^{h,\theta}_{11}) - (h\mu_j^h + E_0)|\leq C_0 h^2  \fin $

\section{Asymptotic of the first band}

For any real Lipschitz $\Gamma $ periodic function
$\phi ,$ and for any $u\in W^2(\bT^n;\bC^2),$ we have the identity
\begin{equation}\label{AgId}
\left . \begin{array}{l}
\MR{Re} \left ( < P^h(x,h(D-\theta))u \Bar e^{2\phi /h} u >_{L^2(\bT^n;
\bC^2)} \right )  =\\
\displaystyle{\sum_{k=1}^{n} } h^2\| ((D_{x_k} -\theta_k)I -A_k)e^{\phi
/h}u\|^{2}_{L^2(\bT^n; \bC^2)}\\
 \quad + < (\widetilde{V}-|\nabla \phi |^2 I +h^2 R)u \Bar e^{2\phi /h} u
>_{L^2(\bT^n; \bC^2)} .
\end{array}
\right \}
\end{equation}
This identity enables us to apply the method used in \cite{He-Sj-1},
(see also \cite{He} and \cite{Ou}).
We define the Agmon \cite{Ag}   distance on $\bR^n$
\begin{equation}\label{AgM}
d(y,x)=\inf_{\gamma } \int_0^1\sqrt{v(\gamma (t))-|w(\gamma (t))| -E_0}
|\, \dot{\gamma} (t)|dt \; ,
\end{equation}
the inf is taken among paths such that $\gamma (0)=y$ and $\gamma (1)=x.$ \\
For common  properties of the Agmon distance, one can see for example
\cite{Hi-Si}.

We will use that, for any fixed $y\in \bR ^n,$
the function $d(y,x)$ is a Lipschitz function on $\bR^n$ and
$|\nabla _x d(y,x)|^2 \leq v(x)-|w(x)|- E_0$ almost everywhere on $\bR^n.$

Using that the zeros of $v(x) - w(x) -E_0$ are the elements of $\Gamma $
and are non degenerate,   we get that
the real function $d_0(x)=d(0,x)$ satisfies, (see \cite{He-Sj-1}),  $\Di
|\nabla d_0(x)|^2=v(x)-|w(x)|- E_0$
in a neighbourhood
of $0.$


We summarize the properties of the Agmon distance we will need:
\begin{equation}\label{Ag}
\left .
\begin{array}{ll}
i) &
\exists \,  R_0>0\ \MR{s. t.}\ d_0(x) \in C^\infty (B_0(R_0))\\
ii) & |\nabla d_0(x)|^2=v(x)-|w(x)|-E_0\; ,\ \forall x\in B_0(R_0)\\
iii) & |\nabla d_0(x)|^2\leq v(x)-|w(x)|-E_0\\
iv) &  |\nabla d_\Gamma (x)|^2\leq v(x)-|w(x)|-E_0
\end{array}
\right \}
\end{equation}
where $d_0(x)=d(0,x),\ B_0(r)=\{ x\in \bR^n ;\ d_0(x)<r\}  $

\No
and $\Di d_\Gamma (x)=d( \Gamma ,x )=\min_{\omega \in \Gamma } d(\omega ,x).$

The least Agmon distance in $\Gamma $ is
\begin{equation}\label{defS}
S_0=\inf_{1\leq k\leq n} d_0( \beta_k) = \inf_{\rho \neq \omega ,\ (\omega
,\rho )\in \Gamma^2} d(\omega ,\rho )\; .
\end{equation}
The Agmon distance  on $\bT^n,\ d^{\bT^n}( . , .),$ is defined by its
$\Gamma$-periodic extension
on $(\bR^n)^2$
\[   d^{\bT^n}( y , x)=\min_{\omega \in \Gamma } d(y,x + \omega ).\]
  Then
\begin{equation}\label{defST}
 \frac{S_0}{2} =\sup_{r} \{ r>0\ s.t.\ \{ x\in \bT^n;\ d^{\bT^n}
(x_0,x)<r\} \ \MR{is}\ \MR{simply\ connected} \}  \; ,
\end{equation}
where $x_0$ is the single point in $\bT^n$ such that $\Di
v(x_0)-|w(x_0)|=E_0.$
The $\Gamma $-periodic function on $\bR^n,\ d_\Gamma (x)$ is the one
corresponding to
the extension of
$d^{\bT^n}(x_0,x).$

If $\lambda^{h,\theta}$ is an eigenvalue of $H^{h,\theta }$ and if
$u^{h,\theta}$ is an associated
eigenfunction, then by (\ref{AgId})  one gets as in the scalar case
considered in \cite{He-Sj-1}) and
\cite{He-Sj-2},
\begin{equation}\label{AgIdE}
\left . \begin{array}{l}
\Di{\sum_{k=1}^{n}} h^2\| ((D_{x_k} -\theta_k)I -A_k)e^{\phi
/h}u^{h,\theta}\|^{2}_{L^2(\bT^n; \bC^2)} \\
\quad + < [\widetilde{V}-|\nabla \phi |^2 I +h^2 R - \lambda^{h,\theta}
I]_+ u^{h,\theta} \Bar e^{2\phi /h} u^{h,\theta} >_{L^2(\bT^n; \bC^2)}\\
= < [\widetilde{V}-|\nabla \phi |^2 I +h^2 R - \lambda^{h,\theta}
I]_-u^{h,\theta} \Bar e^{2\phi /h} u^{h,\theta} >_{L^2(\bT^n; \bC^2)} ,
\end{array} \right \}
\end{equation}
so, when $\phi (x)=d^{\bT^n}(x_0,x),$  necessarily $\lambda^{h,\theta} -
E_0 +\bbO (h^2) >0,$\\
and  if $h/C<\lambda^{h,\theta} - E_0<hC,$
 then   $u^{h,\theta}$ is localized in energy near $x_0,$  for any
$\eta \in ]0,1[,\ \exists C_\eta >0$ such that
\begin{equation}\label{AgIdd}
\left . \begin{array}{l}
\Di{\sum_{k=1}^{n}} h^2\| ((D_{x_k} -\theta_k)I -A_k)e^{\eta
d^{\bT^n}(x_0,x) /h}u^{h,\theta}\|^{2}_{L^2(\bT^n; \bC^2)} \\
\quad + (1-\eta^2 )< (v -|w| -E_0)  u^{h,\theta} \Bar e^{2\eta
d^{\bT^n}(x_0,x) /h} u^{h,\theta} >_{L^2(\bT^n; \bC^2)}\\
 \leq  hC_\eta  \int_{\{ x\in \bT^n;\ d^{\bT^n}(x_0,x)<\sqrt{h} C\} }
 |u^{h,\theta}(x) |^{2} dx .
\end{array} \right \}
\end{equation}
\indent Let $\Omega \subset \bT^n$ an open and simply connected set with
smooth boundary satisfying, for some $\eta ,\ 0<\eta <S_0/2,$
\begin{equation}\label{Omeg}
\{ x\in \bT^n ;\ d^{\bT^n}(x_0,x)< \frac{S_0-\eta }{2}  \} \, \subset \,
\Omega \, \subset \,  \{ x\in \bT^n ;\ d^{\bT^n}(x_0, x)< S_0/2 \}
\end{equation}
Let $H^{h}_{\Omega} $ be the selfadjoint operator
on $L^2(\Omega ; \bC^2)$ associated with $P^h(x,hD)$ with Dirichlet
boundary condition.
We denote by $(\lambda_j(H^{h}_{\Omega}))_{j\in  \bN^\star }$
the increasing sequence of eigenvalues of $H^{h}_{\Omega}.$
 Using the method of \cite{He-Sj-1}, we get easily
the following results
\begin{theorem}\label{compareD}
For any $\eta , \ 0<\eta <S_0/2,$
there exist $h_0>0$ and $ N_0>1$   such that,
 if $0<h<h_0$ and $j\leq N_0,$ then
\begin{equation}\label{compareD1}
 \forall \theta \in \bB^n\, ,\quad
0\,  <\, \lambda_j(H^{h}_{\Omega})\,  - \,  \lambda_{j}^{h, \theta}
\, \leq \, Ce^{-(S_0-\eta )/(2h)}
\, ;
\end{equation}
 so the length of the band $b_j^h$ satisfies
  $|b_j^h| \leq Ce^{-(S_0-\eta )/(2h)}.$

For the first band, we have the following  improvement
\begin{equation}\label{compareD2}
|b_1^h| \leq Ce^{-(S_0-\eta )/h}.
\end{equation}

\end{theorem}
{\it Sketch of the proof. }

As $\Omega $ is simply connected and the one form
$\theta dx$ is closed, there exists a  smooth real function
$\psi_\theta (x)$ on $\overline{\Omega }$ such that
$e^{-i\psi_\theta (x)} P^h(x,hD)e^{i\psi_\theta (x)} =
P^h(x,h(D-\theta )): $ the Dirichlet operators on $\Omega ,\ H^{h}_{\Omega}$
and $H^{h, \theta }_{\Omega}$ associated to
$P^h(x,hD)$ and $P^h(x,h(D-\theta ))$ are gauge equivalent, so they have
the same spectrum.

Therefore the min-max principle says that
\[ 0  < \lambda_j(H^{h, \theta }_{\Omega})  -   \lambda_{j}^{h, \theta}
=\lambda_j(H^{h}_{\Omega})\,  - \,  \lambda_{j}^{h, \theta}.\]
But the exponential decay of the eigenfunction $\varphi_{j}^{h,\theta}(x)$
associated with
$\lambda_{j}^{h, \theta},$ given by (\ref{AgIdd})  implies that
\[  \| (P^h(x,h(D-\theta )) -  \lambda_{j}^{h, \theta}) \chi
\varphi_{j}^{h,\theta}(x)\|_{L^2(\Omega ; \bC^2)} \leq Ce^{-(S_0 -\eta
+\epsilon )/(2h)}, \]
for some  $\epsilon >0 ,$ and for
a smooth cut-off function $\chi $ supported in $\Omega $ and
$\chi (x)=1$ if $d^{\bT^n}(x_0,x)\leq (S_0-\eta +\epsilon )/2.$

So $distance(\lambda_{j}^{h, \theta} , \sp (H^{h}_{\Omega}))
\leq Ce^{-(S_0 -\eta +\epsilon )/(2h)}.$  \\
This achieves the proof of (\ref{compareD1}).

 Let us denote $E^{h,\theta},$ (respectively $ E^{h,\Omega}),$
the first eigenvalue $\lambda_{1}^{h, \theta} ,$ (respectively $ \
\lambda_1(H^{h}_{\Omega})),$ and
$\varphi ^{h,\theta}(x),$ (respectively $\varphi ^{h,\Omega}(x)),$ the
associated normalized eigenfunctions.
Let $\chi $ be a cut-off function satisfying the same properties as before.
Then \\
$\Di P^h(x,h(D-\theta ))(e^{-i\psi_\theta (x)} \chi (x)
\varphi ^{h,\Omega}(x))= \lambda_1(H^{h}_{\Omega}) e^{-i\psi_\theta (x)}
\chi (x)
\varphi ^{h,\Omega}(x)\, +\; e^{-i\psi_\theta (x)} r_0^h (x)$\\
with, thanks to  the same identity as (\ref{AgIdd})
for Dirichlet problem on $\Omega ,$
 \[  \| r_0^h \|_{L^2(\bT^n ;\bC^2)} \leq Ce^{-\frac{S_0-\eta}{2h} }.
\]
 The same argument used in \cite{He-Sj-1}, (see also \cite{He}),
 gives this estimate
 \[ | E^{h,\theta} - E^{h,\Omega} - < r_0^h \Bar
  \chi
\varphi ^{h,\Omega} >_{L^2(\bT^n ;\bC^2)} |\leq Ce^{-(S_0 -\eta )/h}.
\]
As  $\tau_{h}=< r_0^h \Bar
  \chi
\varphi ^{h,\Omega} >_{L^2(\bT^n ;\bC^2)}
$ does not depend on $\theta ,$
so
\[ \forall \theta \in \bB^n\; ,\quad | E^{h,\theta} - E^{h,\Omega} -
\tau_{h} | \leq Ce^{-(S_0 -\eta )/h}
\]
this estimate ends the proof of  (\ref{compareD2}) $\Box $

 As  for the tunnel effect in \cite{He-Sj-1} and \cite{Si-2},
we have an accuracy
estimate for the first band, like the scalar case
in \cite{Si-3} and in \cite{Ou} (see also \cite{He}).

\begin{theorem}\label{1band}
There exists $h_0> 0$    such that,
 if $0<h<h_0$  then
\[ |b_1^h|
\leq C e^{-S_0/h}
.\]
\end{theorem}
{\it Sketch of the proof. } Instead of comparing $H^{h, \theta }$ with
an operator defined in a subset of $\bT^n,$ we have to work  on the
universal cover $\bR^n$ of $\bT^n.$

We take  $\Omega \subset \bR^n$ an open and simply connected set with
smooth boundary satisfying, for some $\eta_0 ,\ 0<\eta_0<\eta_1 <S_0/2,$
\begin{equation}\label{OmegR}
B_0((S_0+\eta_0 )/2) \, \subset \,
\Omega \, \subset \,  B_0((S_0+\eta_1 )/2).
\end{equation}
So $\Omega $ contains the Wigner set $\bW^n,$ more precisely
\[ \bW^n \, \subset \, \Omega \, \subset \, 2 \bW^n\quad \MR{and} \quad
\Omega \, \cap \, \Gamma =\{ 0\} .\]
\indent
We let also denote  $H^{h}_{\Omega} $  the Dirichlet operator
on $L^2(\Omega ; \bC^2)$ associated with $P^h(x,hD),$
and  $E^{h}_{\Omega}$ its first eigenvalue.
The associated eigenfunction is also denoted by
$\varphi ^{h,\Omega}(x).$

In the same way as to get (\ref{AgIdd}), we have

\begin{equation}\label{AgD1}
\sum_{k=1}^{n} h^2\| (D_{x_k} I -A_k)e^{ d_0(x) /h} \varphi
^{h,\Omega}\|^{2}_{L^2(\Omega ; \bC^2)}
 \leq  hC \int_{ B_0(\sqrt{h} C) }
 |\varphi ^{h,\Omega}(x) |^{2} dx ,
\end{equation}
then the Poincar\'e estimate gives
\begin{equation}\label{AgD0}
\int_\Omega e^{ 2d_0(x) /h} |\varphi ^{h,\Omega} (x)|^2dx
\leq  h^{-1} C \int_{ B_0(\sqrt{h} C) }
 |\varphi ^{h,\Omega}(x) |^{2} dx .
\end{equation}

Let $\chi $ a smooth cut-off function satisfying
\[ \chi (x)\, =\, 1\ \MR{if}\ d_0(x)\leq (S_0+\eta_0 )/2
\quad \MR{and}\quad \chi (x)\, =\, 0 \ \MR{if}\ x\notin \Omega .
\]
Then the function
\[ \varphi ^{h, \theta}(x)\, =\, \sum_{\omega \in \Gamma } e^{i\theta
(x-\omega )} \chi (x-\omega ) \varphi ^{h,\Omega}(x-\omega )
\]
is $\Gamma$-periodic and satisfies
\begin{equation}\label{Int1}
\left . \begin{array}{l}
 (P^h(x, h(D-\theta )) - E^{h}_{\Omega})\varphi ^{h,
\theta}(x)=r^{h,\theta} \ \MR{and} \\
\| r^{h,\theta}\|_{L^2(\bW^n; \bC^2)} \,
 \leq \, Ce^{-(S_0+\eta_0)/(2h)}\| \varphi ^{h, \theta}\|_{L^2(\bW^n; \bC^2)}
\end{array} \right \}
\end{equation}
and \\
$\Di < r^{h,\theta}\, |\, \varphi ^{h, \theta}>_{L^2(\bW^n; \bC^2)} \, =$
\[ \sum_{\omega , \rho \in \Gamma_0} e^{i\theta (\rho - \omega )}
\int_{\bW^n} ([P^h(x,hD); \chi ] \varphi ^{h,\Omega})(x-\omega).
\overline{ (\chi \varphi ^{h,\Omega})} (x-\rho ) dx\]
with $\Gamma_0=\{ 0, \pm \beta_1,\ldots , \pm \beta_n\} $ and
\[  [P^h(x,hD); \chi ] =-2h^2i\sum_{k=1}^{n}
\partial_{x_k}  \chi  (D_{x_k}I-A_k) \,  -\, h^2\Delta \chi I.\]
So
\begin{equation}\label{Int2}
\left |  \frac{1}{\| \varphi ^{h, \theta}\|^{2}_{L^2(\bW^n; \bC^2)} }
< r^{h,\theta}\, |\, \varphi ^{h, \theta}>_{L^2(\bW^n; \bC^2)}
 \right |
\, \leq \, C e^{-S_0/h} .
\end{equation}
The proof comes easily from (\ref{AgD1}) and (\ref{AgD0}) as
in \cite{Ou} or in \cite{He}.

Using the same argument of \cite{He-Sj-1} as in the proof of
(\ref{compareD2}), we
get that
\begin{equation}\label{Int3}
| E^{h, \theta} \, -\, E^{h}_{\Omega} -\tau^{h,\theta} |\, \leq \,
Ce^{-(S_0+\eta_0)/h}) \, ,
\end{equation}
with $\Di \tau^{h,\theta}=  \frac{1}{\| \varphi ^{h,
\theta}\|^{2}_{L^2(\bW^n; \bC^2)} }
< r^{h,\theta}\, |\, \varphi ^{h, \theta}>_{L^2(\bW^n; \bC^2)} .$

\No
 Theorem \ref{1band}  follows from (\ref{Int2}) and (\ref{Int3})
$\Box $

\section{B.K.W. method for the Dirichlet ground state}
Let $ \Omega $  be an open set satisfying (\ref{Omeg}).  more precisely
 $\Omega \subset \bR^n$ an open, bounded and simply connected  set with
smooth boundary satisfying, for some\\
 $\eta_1$ and $\eta_2 ,\ 0<\eta_1<\eta_2 <S_0/2,$
\begin{equation}\label{OmegB}
\{ x\in \bR^n ;\ d_0(x)< \frac{S_0-\eta_2 }{2}  \} \, \subset \,
\Omega \, \subset \,  \{ x\in \bR^n ;\ d_0(x)< \frac{S_0-\eta_1 }{2}  \}
\end{equation}

\begin{theorem}\label{aedir}
The first eigenvalue $E^{h,\Omega} =\lambda_1(H^h_\Omega )$ of the
Dirichlet operator $H^h_\Omega $ admits an asymptotic expansion of the
form
\[  E^{h,\Omega}  \simeq  \sum_{j=0}^{\infty} h^j e_j \ ,
\]
 and if $S_0-\eta_1$ is small enough,  the associated  eigenfunction
$\varphi^{h,\Omega }$ has also an asymptotic expansion of the form
\[ \varphi^{h,\Omega } = e^{-\phi/h}(f_h^+, f_h^-)\ , \quad f_h^\pm \simeq
\sum_{j=0}^{\infty} h^j f_j^\pm \, ,\quad (f_0^-=0)\, .
\]
As usual
\begin{equation}\label{e012}
e_0=E_0\, ,\quad e_1=\tau_1\, ,\quad
e_2=r_{11}(0)+\sum_{k=1}^{n}|a_{k,11}(0)|^2\, ,
\end{equation}
and
 $\phi$ is the real function   satisfying the eikonal equation
\begin{equation}\label{eik}|\nabla \phi (x)|^2 =v(x) -|w(x)| - E_0 \,
,\end{equation}
equal to $d(x)$ in a neighbourhood of $0.$\\
$(r_{11}$ and the $a_{k,11}$ are defined by (\ref{type}) and (\ref{coeff}).
$\ E_0$ and $\tau_1$ are defined by (\ref{carre}) and (\ref{H1b})).
\end{theorem}
\demo
When  the gauge potential matrix is identified with the one form
\[ Adx\, =\, \sum_{k=1}^{n}A_k(x)dx_k\, ,\]
 its curvature form appears to be the related magnetic field $B=dA +
A\wedge A:$
\[  B= \sum_{1\leq j<k\leq n}( \partial_{x_j}A_k(x) -
 \partial_{x_k}A_j(x))dx_j\wedge dx_k \, +\,
 \sum_{1\leq j<k\leq n}( A_j(x)A_k(x) -A_k(x)A_j(x))dx_j\wedge dx_k.\]
For our purpose, only the vector magnetic potential
$a_{. ,11}$ is significant. We will work with Coulomb vector gauge
$a_{. ,11}:$
\begin{equation}\label{zeroMag}
div(a_{. , 11}(x))= \sum_{k=1}^{n} \partial_{x_k} a_{k,11}(x) \, =\, 0 \, .
\end{equation}
It is feasible thanks to the existence of a smooth real and $\Gamma $
periodic function
$\psi (x)$ such that $\Delta \psi (x) = div(a_{., 11}(x)) .$\\
Let $\cO $ be  any open set of $\bR^n $ (or one can take also $\cO =\bT^n).$
 Conjugation of  $P^h(x,hD)$ by the unitary operator $J_\psi $
on  $L^2(\cO ; \bC^2) $ :
\begin{equation}\label{Jpsi} 
J_\psi  \, =\, \left ( \begin{array}{cc}
e^{i\psi} & 0 \\ 0 & e^{-i\psi} 
\end{array} \right ) , 
\end{equation} 
 leads to changing
$a_{., 11}(x)$ for $a_{., 11}(x) -\nabla \psi (x)$ and $a_{., 21}(x)$ for $e^{2i\psi (x)}a_{., 21}(x)$; the new $a_{.,22}(x)$ is equal
to minus
  the new $a_{., 11}(x),$  
and   
the new $ a_{., 12}(x)$ remains the conjugate 
of the new $a_{., 21}(x).$
So by  (\ref{Atype1}) we have
\begin{equation}\label{NewA}
 \left .
\begin{array}{l}
a_{., 11}   = \beta \nabla \alpha - \alpha \nabla \beta
+\delta \nabla \rho
- \rho \nabla \delta
  - \nabla \psi  \, ,\\
a_{., 21}  =
e^{2i\psi } [ (\rho \nabla \beta - \beta \nabla \rho
+\delta \nabla \alpha
- \alpha \nabla \delta )
+ i( \alpha \nabla \rho - \rho \nabla \alpha
+\delta \nabla \beta
- \beta \nabla \delta )]\\
a_{., 22} = - a_{., 11}\, ,\quad a_{., 12} =\overline{a_{., 21}}\\
\Delta \psi  =div (\beta \nabla \alpha - \alpha \nabla \beta
+\delta \nabla \rho
- \rho \nabla \delta  )
\end{array} \right \}
\end{equation}

Let us write
\begin{equation}\label{transf}
 e^{-\phi /h} P^h(x,hD)(e^{-\phi /h} f_h)=W_0(x) f_h \, +\, hW_1(x,D)f_h
\, +\, h^2W_2(x,D)f_h
\end{equation}
with
\[  \begin{array}{l}
W_0(x)=V(x) -|\nabla \phi (x)|^2I\\
W_1(x,D) =  \Delta \phi I +2i\Di{\sum_{k=1}^{n}}
\partial_{x_k} \phi (D_{x_k} I-A_k)\\
W_2(x,D) = \Di{\sum_{k=1}^{n}}
(D_{x_k} I-A_k)^2\, +\, R(x).
\end{array}
\]
So
\[ W_1(x,D) =\left (   \begin{array}{cc}
\Delta \phi -2i\nabla \phi .(i\nabla +a_{. , 11}) & -2i \nabla \phi
.\overline{a_{. , 21}}\\
-2i \nabla \phi . a_{. , 21} & \Delta \phi -2i\nabla \phi .(i\nabla -a_{.
, 11})
\end{array} \right ) \, ,\]
and
\[ W_2(x,D) =\left (   \begin{array}{cc}
 (i\nabla +a_{., 11})^2 + r_{11} & (i\nabla  +a_{., 11}).\overline{a_{.,
21}} +
\overline{a_{., 21}}. (i\nabla  - a_{., 11}) + r_{12}\\
a_{., 21}.(i\nabla +a_{., 11}) + (i\nabla +a_{., 11}).a_{., 21} + r_{21}
& (i\nabla - a_{., 11})^2 +r_{22}
\end{array} \right ) \, .\]
We look for an eigenvalue $\Di E^h\simeq \sum_{j=0}^{\infty} h^j e_j$\\
and an associated eigenfunction $f_h
\simeq \sum_{j=0}^{\infty}h^j f_j,$ so
\[  e^{\phi /h}(P^h(x,hD)-E^hI)(e^{-\phi /h} f_h)
\simeq \sum_{j=0}^{\infty}h^j \kappa_j
\]
with
\[ \begin{array}{l}
\kappa_0(x)=(W_0(x) -e_0I)f_0(x)\\
\kappa_1(x)=(W_1(x,D) -e_1I)f_0(x) + (W_0(x) -e_0I)f_1(x)\\
\kappa_2(x)=(W_2(x,D) -e_2I)f_0(x) + (W_1(x,D) -e_1I)f_1(x)+ (W_0(x)
-e_0I)f_2(x)\\
\kappa_j(x)= -e_jf_0(x) - \sum_{\ell =1}^{j-3} e_{j-\ell}f_\ell (x) +
(W_2(x,D) -e_2I) f_{j-2}(x) \\
\quad \quad + \quad (W_1(x,D) -e_1I) f_{j-1}(x)+ (W_0(x) -e_0I)f_j(x),\
(j>2).
\end{array}
\]
We recall that  $f_j=(f_j^+ ,f_j^-)$ and we want that
$\kappa_j(x)=0,\ \forall j.$

\textbf{1) Term of order 0}\\
As $\Di \kappa_0(x)=0 \, \Leftrightarrow \, \left \{ \begin{array}{l}
(- |\nabla \phi (x)|^2 + v(x) -|w(x)|- e_0) f_0^+ (x)  =0 \\
(- |\nabla \phi (x)|^2 + v(x) +|w(x)|- e_0) f_0^- (x)=0\, ,
\end{array} \right . $\\
choosing $\phi $ satisfying (\ref{eik}), then

$e_0=E_0$ and
$- |\nabla \phi (x)|^2 + v(x) +|w(x)|- e_0=2|w(x)|>0$
implies that
\begin{equation}\label{azero0}
f_0^- (x)\, =\, 0\, , \quad e_0=E_0 \quad \MR{and}\quad W_0(x) -e_0I=\left
( \begin{array}{cc} 0 & 0\\ 0 & 2|w(x)| \end{array} \right )\, .
\end{equation}

\textbf{2) Term of order 1.} \\
The components of  $\kappa_1(x)=(\kappa_1^+(x), \kappa_1^-(x))$ become
\[ \begin{array}{l}  \kappa_1^+(x)=
(\Delta \phi (x) -e_1 - 2i a_{., 11}(x).\nabla \phi (x))f_0^+(x)  +2\nabla
\phi (x).\nabla  f_0^+(x)   \\
\kappa_1^-(x)=2|w(x)| f_1^-(x)
 \,-\, 2i(a_{.,21}(x).\nabla \phi (x)) f_0^+(x)\, ,
\end{array}
\]
As $|\nabla \phi (x)|$ has a simple zero
at $x_0=0,$ the  equation  $\kappa_1^+(x)=0$
can be solved only when $e_1=\Delta \phi (0).$
In this case there exists a unique function $f_0^+(x)$
such that $f_0^+(0)=1$ and $\kappa_1^+(x)=0.$
We can conclude that the study of the term of order 1
 leads to
\begin{equation}\label{azero00}
\left .
\begin{array}{l}
e_1=\Delta \phi (0)\, ,\\
 2\nabla \phi (x).\nabla  f_0^+(x) =[e_1 - \Delta \phi (x) + 2 ia_{.,
11}(x).\nabla \phi (x)]f_0^+(x)\, ,\\
f_1^-(x)= \frac{i}{|w(x)|} (\nabla \phi (x).a_{., 21} (x)) f_0^+(x)\, .
\end{array} \right \}
\end{equation}

\textbf{3) Term of order 2.} \\
The components of  $\kappa_2(x)=(\kappa_2^+(x), \kappa_2^-(x))$ become
\begin{equation}\label{t2}
\left .
\begin{array}{l}  \kappa_2^+(x)=(\, (i\nabla + a_{., 11})^2  ++r_{11}(x)-
e_2)f_0^+(x)
 \\
\quad + (\Delta \phi (x) -e_1 - 2 ia_{., 11}(x).\nabla \phi (x))f_1^+(x) 
+2\nabla \phi (x).\nabla  f_1^+(x) \\
\quad  -\, 2i(\overline{a_{.,21}(x)}.\nabla  \phi (x)) f_1^-(x) \\
\kappa_2^-(x)=2|w(x)| f_2^-(x)
 - 2i(a_{.,21}(x).\nabla \phi (x)) f_1^+(x)
  \\
\quad   +2\nabla \phi (x).\nabla f_1^- (x)  + 2 ia_{., 11}(x).\nabla \phi
(x)f_1^- (x)
+ 2ia_{.,21} (x).\nabla f_0^+ (x)\\
\quad  +\, i(div(a_{.,21} (x))f_0^+ (x)
+ r_{21}(x)f_0^+ (x)
\, ,
\end{array} \right \}
\end{equation}
The unknown function $f_1^+(x)$ must give
$\kappa_2^+(x)=0$ in (\ref{t2}). This equation, with
the initial condition $f_1^+(0)=0,$
can be solved only if
\[ e_2= (i\nabla +a_{., 11})^2 f_0^+(0) +r_{11}(0).
\]
(We used that $f_0^+(0)=1).$
So $\kappa_2=0$ implies
\[
\begin{array}{l}
e_2\, =\, (i\nabla +a_{., 11})^2 f_0^+(0) +r_{11}(0)\, ,\\
2\nabla \phi (x).\nabla  f_1^+(x)\, =\,  -
 (\Delta \phi (x) -e_1 - 2 ia_{., 11}(x).\nabla \phi (x))f_1^+(x)\\
\quad  -
(\,  (i\nabla +a_{., 11})^2
  + e_2  - r_{11}(x))f_0^+(x)
+2i(a_{.,12}(x).\nabla  \phi (x)) f_1^-(x)\\
f_2^-(x)\, =\, \frac{1}{2|w(x)|}
[ 2i(a_{.,21}(x).\nabla \phi (x)) f_1^+(x)
 -2\nabla \phi (x).\nabla f_1^- (x)
-2i(a_{.,11}(x) .\nabla \phi (x))   f_1^- (x)\\
\quad \quad
-2ia_{.,21} (x).\nabla f_0^+ (x) -i(div(a_{.,21} (x))f_0^+ (x)
- r_{21}(x)f_0^+ (x)]\, .
\end{array}
\]
\indent
\textbf{4) Term of order j}$\mathbf{>2.}$  \\
We assume that $e_\ell $ for $\ell =0, 1,\ldots ,j-1,$
 the functions $f_\ell^\pm (x)$
for $\ell =0, 1,\ldots ,j-2,$ and the one
$f_{j-1}^{-} (x)$ are well-known,
$f_\ell^+ (0)=0$ when $0<\ell <j-1.$

The equation $\kappa_j^+=0$ becomes
\begin{equation}\label{kapP}
\left . \begin{array}{l}
2\nabla \phi (x).\nabla  f_{j-1}^{+}(x)\, +\,
(\Delta \phi (x) -e_1 - 2 ia_{., 11}(x).\nabla \phi (x))f_{j-1}^{+}(x)\,
=\, \\
2i(\overline{a_{.,21}(x)}.\nabla  \phi (x)) f_{j-1}^{-}(x)
+\Di{\sum_{\ell =0}^{j-3}} e_{j-\ell } f_\ell^+(x)\\
-(\, (i\nabla + a_{., 11})^2  - e_2
+ r_{11}(x)\, )f_{j-2}^{+}(x) \\
-(\,  r_{12}(x) +i div(\overline{a_{21}})\, )f_{j-2}^{-}(x)
 -2i \overline{a_{21}}.\nabla f_{j-2}^{-}(x)
\end{array} \right \}
\end{equation}
 This equation has a unique solution $f_{j-1}^{+}(x)$ such that
$f_{j-1}^{+}(0)=0$ iff
\begin{equation}\label{ej}
\left . \begin{array}{l}
e_j\, =\, (\,  (i\nabla + a_{., 11})^2 \, +\, r_{11}(0) \, )f_{j-2}^{+}(0)
+  r_{12}(0))f_{j-2}^{-}(0) \\
-2i\overline{a_{. ,21}(0)}.\nabla )f_{j-2}^{-}(0)
-\sum_{\ell =0}^{j-3} e_{j-\ell} f_{\ell}^{+}(0) \, .
\end{array} \right \}
\end{equation}
The equation $\kappa_j^-=0$ gives
\[
f_j^-(x)\, =\, \frac{1}{2|w(x)|}\times \]
$\Di
[-2\nabla \phi (x).\nabla f_{j-1}^{-}(x) +   (e_1 -2i a_{.,11}(x).\nabla
\phi (x))f_{j-1}^{-}(x)
+2i  a_{.,21}(x).\nabla \phi (x)f_{j-1}^{+}(x) $\\
$\Di
+(e_2 -(i\nabla -a_{., 11})^2)f_{j-2}^{-} (x)
- i div( a_{.,21}(x)) f_{j-2}^{+} (x) -2i  a_{.,21}(x).\nabla  f_{j-2}^{+}
(x)
+\sum_{\ell =0}^{j-3} e_{j-\ell} f_{\ell}^{-}(x)]
$

\indent
\textbf{5) End of the proof.}\\
Let $\chi (x)$ be a cut-off function equal to $1$ in a neighbourhood of $0$
and supported in $\Omega .$
Then taking $\chi (x)f_j(x)$ instead  of $f_j(x),$
 we get a function $\varphi^{h, \Omega}$  satisfying
Dirichlet boundary condition  and Theorem \ref{aedir}.
The self-adjointness of the related operator ensures that the computed
sequence
$(e_j)$ is real $\Box $

\section{Sharp asymptotic for the  width of the first band}

Returning to the proof of Theorem \ref{1band}, we have to study carefully the
$\tau^{h,\theta}$ defined in (\ref{Int3}), using the method of \cite{He-Sj-1}
performed in \cite{He}
and \cite{Ou}.

Using  (\ref{Int1})--(\ref{Int3}), we have
\begin{equation}\label{Split1}
\tau^{h,\theta}=\sum_{\omega \in \Gamma^+_0 } \left ( e^{-i\theta \omega}
(\rho^{+}_{\omega} + \rho^{-}_{\omega}) + e^{i\theta \omega}
\overline{(\rho^{+}_{\omega} + \rho^{-}_{\omega}) } \right ) \, ,
\end{equation}
with $\Gamma^+_0 =\{  \beta_1,\ldots ,  \beta_n\}  \, ,$
\[ \rho^{+}_{\omega}= \frac{1}{ \| \varphi ^{h, \theta} \|^{2}_{L^2(\bW^n;
\bC^2)} }
\int_{\bW^n} ([P^h(x,hD); \chi ] \varphi ^{h,\Omega})(x - \omega).
\overline{ (\chi \varphi ^{h,\Omega})} (x ) dx  \]
and $\Di \rho^{-}_{\omega}= \frac{1}{ \| \varphi ^{h, \theta}
\|^{2}_{L^2(\bW^n; \bC^2)}}
\int_{\bW^n} ([P^h(x,hD); \chi ] \varphi ^{h,\Omega})(x ).
\overline{ (\chi \varphi ^{h,\Omega})} (x + \omega) dx \, .$\\
We get from the formula of $[P^h(x,hD); \chi ] $ and from the estimate
(\ref{AgD0})
that
\begin{equation}\label{Split2}
\rho^{+}_{\omega}= - \frac{h^2}{\| \varphi ^{h, \theta} \|^{2}_{L^2(\bW^n;
\bC^2)} }
\int_{\bW^n} (\nabla \chi (x - \omega)\nabla \varphi ^{h,\Omega})(x -
\omega)).
\overline{ (\chi \varphi ^{h,\Omega})} (x ) dx  \, +
\end{equation}
\[+ \, \textbf{O} (h e^{-S_0/h})\]
and
\[ \rho^{-}_{\omega}= -\frac{h^2}{ \| \varphi ^{h, \theta}
\|^{2}_{L^2(\bW^n; \bC^2)}}
\int_{\bW^n} (\nabla \chi (x )\nabla \varphi ^{h,\Omega})(x )).
\overline{ (\chi \varphi ^{h,\Omega})} (x + \omega) dx
 \, +\, \textbf{O} (h e^{-S_0/h}) \, .\]
 \begin{theorem}\label{ThSplit} Under the assumption of Theorem \ref{1band},
 if for any $\omega \in \{  \pm \beta_1,\ldots ,  \pm  \beta_n\}  $ such
that the
Agmon distance in $\bR^n$ between $0$ and $\omega $
is the least one, (i.e. $d(0,\omega )=S_0 ),$ there exists one or a finite
number
of minimal geodesics joining $0$ and $\omega ,$ then
there exists $\eta_0>0$ and $h_0>0$ such that
\[ b_1^h=\eta_0h^{1/2}e^{-S_0/h}\left ( 1\, +\; \textbf{O}(h^{1/2}) \right
)\; ,\quad
\forall h \in ]0,h_0[\, .\]
\end{theorem}
 {\it Sketch of the proof. }
Following the proof of splitting in \cite{He-Sj-1}
and \cite{He}, in (\ref{Split2}) we can change
$\bW^n$ into $\bW^n\cap \cO ,$ where $\cO $ is any
neighbourhood of the minimal geodesics between
$0$ and the $\pm \beta_k,$ such that
$d(x)=d(0,x)\in C^\infty (\cO ).$
 In this case the B.K.W. method is valid in
$\bW^n\cap \cO .$ If $\varphi ^{h}_{B.K.W.}$ is the
B.K.W. approximation of $\varphi ^{h,\Omega}$ in $ \bW^n\cap \cO ,$
then, thanks to (\ref{AgId}), for any $p_0>0$ there exists $C_{p_0}$ such
that
\\
$\Di
h\sum_{k=1}^{n}\| (D_{x_k}I-A_k)e^{d(x)/h} \chi_0
(\varphi ^{h,\Omega}-\varphi ^{h}_{B.K.W.})\|^{2}_{L^2(\bW^n; \bC^2)}$
\begin{equation}\label{compare0}
+\| e^{d(x)/h} \chi_0
(\varphi ^{h,\Omega}-\varphi ^{h}_{B.K.W.})\|^{2}_{L^2(\bW^n; \bC^2)}
\, \leq \, h^{p_0}C_{p_0}\, ,
\end{equation}
where $\chi_0$ is a cut-off function supported in $\bW^n\cap \cO $
and equal to $1$ in a neighborhood of the minimal geodesics
between
$0$ and the $\pm \beta_k.$ We have assumed that
$\| \varphi ^{h,\Omega}\|_{L^2(\bW^n; \bC^2)}=1$ and then
$\| \chi_0  \varphi ^{h}_{B.K.W.})\|^{2}_{L^2(\bW^n; \bC^2)}
-1 = \textbf{O}(h^p)$ for any $p>0.$

As  (\ref{Split2}) remains valid if we change $ \varphi ^{h,\Omega}$ into
$\chi_0\varphi ^{h,\Omega},$
the estimate (\ref{compare0}) allows also to change $ \varphi ^{h,\Omega}$
into
$\chi_0\varphi ^{h}_{B.K.W.}.$  As a consequence, Theorem \ref{ThSplit}
follows easily, if in $\bW^n\cap \cO ,\ \chi (x)=\chi_1 (d(x))$
for a decreasing function $\chi_1$ on $[0,+\infty [$ with compact support,
equal to $1$ in a neighborhood of $0.$ In this case (\ref{Split2}) becomes
\begin{equation}\label{Split2B}
\rho^{+}_{\omega}=
\frac{h^{(2-n)/2}}{(2\pi )^{1/2}\prod_{k=1}^{n}\tau_{k}^{1/2} } \times
\end{equation}
$\Di \int_{\bW^n \cap \cO } \chi_{1}^{\prime}(d(x-\omega
))\chi_1(d(x))|\nabla d(x - \omega)|^2 f_{0}^{+}(x-\omega
)f_{0}^{+}(x)e^{-(d(x - \omega)+d(x))/h} dx  \, + \, \textbf{O} (h
e^{-S_0/h})$\\
and
\[   \rho^{-}_{\omega}=
\frac{h^{(2-n)/2}}{(2\pi )^{1/2}\prod_{k=1}^{n}\tau_{k}^{1/2} }
\times \]
$\Di \int_{\bW^n \cap \cO } \chi_{1}^{\prime}(d(x ))\chi_1(d(x+\omega
))|\nabla d(x)|^2 f_{0}^{+}(x )f_{0}^{+}(x+\omega )e^{-(d(x)+d(x+\omega
))/h} dx
 \, +\, \textbf{O} (h e^{-S_0/h}) \, .$

We remind that for any $y$ in a minimal geodesic joining
$0$ to $\pm \beta_k,$ if $y\neq 0$ and $y\neq \pm \beta_k,$ then
 the function  $d(x)+d(x\mp \beta_k),$ when it is restricted to
 any hyper-surface orthogonal to the geodesic through $y,$
has a non degenerate minimum $S_0$ at $y \Box $

\begin{center}\textbf{Acknowledgement}\\
The authors  are grateful to Bernard Helffer for many discussions.
\end{center}

\end{document}